\documentclass[12pt,nofootinbib,preprintnumbers]{revtex4}
\usepackage{latexsym,graphics}
\newcommand{\adotdota}{\left(\frac{a''}{a}\right)}
\newcommand{\SH}{{\cal{H}}}

\newcommand{\amv}[1]{a{}_{-\vec{\mathbf{#1}}}}
\newcommand{\av}[1]{a{}_{\vec{\mathbf{#1}}}}
\newcommand{\avp}[1]{a_{\vec{\mathbf{#1}}'}}
\newcommand{\adv}[1]{a^{\dagger}{}_{\vec{\mathbf{#1}}}}
\newcommand{\admv}[1]{a^{\dagger}{}_{-\vec{\mathbf{#1}}}}
\newcommand{\advp}[1]{a^{\dagger}{}_{\vec{\mathbf{#1}}'}}
\newcommand{\YLI}[2]{Y_{\vec{\mathbf{#1}}}^{(#2)}{}_{i}}

\newcommand{\YSLI}[2]{Y_{\vec{\mathbf{#1}}}^{*(#2)}{}_i}

\newcommand{\abk}{|\vec{\mathbf{k}}|}

\newcommand{\myfigure}[2]{\resizebox{#1}{!}{\includegraphics{#2}}}

\begin{document}
\preprint{EFI-04-27}
\preprint{astro-ph/yymmnn}
\title{Can We See Lorentz-Violating Vector Fields in the CMB?}
\author{Eugene A. Lim}
\affiliation{Department of Astronomy and Astrophysics, Enrico Fermi Institute, and Kavli Institute for Cosmological Physics,  \\
  University of Chicago.}
\email{elim@oddjob.uchicago.edu}

\begin{abstract}
We investigate the perturbation theory of a fixed-norm, timelike Lorentz-violating vector field. After consistently quantizing the vector field to put constraints on its parameters, we compute the primordial spectra of perturbations generated by inflation in the presence of this vector field. We find that its perturbations are sourced by the perturbations of the inflaton; without the inflaton perturbation the vector field perturbations decay away leaving no primordial spectra of perturbations. Since the inflaton perturbation does not have a spin-1 component, the vector field generically does not generate any spin-1 ``vector-type'' perturbations. Nevertheless, it will modify the amplitude of both the spin-0 ``scalar-type'' and spin-2 ``tensor-type'' perturbation spectra, leading to violations of the inflationary consistency relationship. 

\end{abstract}

\maketitle

\section{Introduction}

Lorentz invariance is a cornerstone of both the standard model of particle physics and general relativity, two very successful descriptions of the observable universe. The former describes at the quantum level basic processes involving particles and forces, while the latter describes at the classical level long-range gravitational forces. It is believed that both theories will merge to form a complete quantum theory of gravity at some very high cut-off energy scale like the Planck-scale \cite{Kostelecky:2001xz,Kostelecky:2003fs}.

This cut-off scale marks the point where our old description of nature breaks down, and it is not inconceivable that one of the victims of this breakdown is Lorentz invariance. It is thus interesting to test the robustness of this symmetry at the highest energy scales \cite{Dubovsky:2001hj,Coleman:1998ti,Vankov:2002gt,Kifune:1999ex,Aloisio:2000cm,Bertolami:1999dc,Bertolami:1999da,Alfaro:2002ya,Coleman:1997xq}.

The early universe provides such a laboratory to us. In this paper, we study the effects of such a fixed-norm, constant timelike Lorentz-Violating (LV) vector field (introduced in \cite{Kostelecky:1989jw,Jacobson:2001yj,Jacobson:2004ts}) during inflation and its imprints on the Cosmic Microwave Background (CMB). In order to do this, we extend the background theory discussed in a companion paper \cite{Carroll:2004ai} into the regime of linear perturbations, investigating both the consistency of its quantum field theory and its evolution in an inflationary universe.

Furthermore, in the theory of inflation, very short distance quantum fluctuations are ``blown-up'' by the accelerated expansion, generating a spectrum of primordial super-Hubble size fluctuations. Thus it is reasonable to ask whether the quantum perturbations of this vector field may undergo a similar procedure to leave us with a classical spectrum of perturbations. Combined with the fact that vector perturbations can source both B-type and E-type polarization modes \cite{Hu:1997hp}, such a primordial spectrum of vector modes could conceivably leave strong signatures on the CMB. 

The question now becomes whether perturbations of this vector field will be frozen-out during the inflationary epoch, akin to that of the inflaton and gravitational wave perturbations. As we will see in Section (\ref{SectiondS}), the answer to this question is no.

Nevertheless, while we are thwarted in our attempt to directly see this vector field, it does leave other more subtle effects on the CMB. We investigate these effects in Section (\ref{SectionInf}), where we show that the vector field modifies both the spin-0 ``scalar'' and spin-2 ``tensor'' perturbation spectrum in possibly detectable ways. It follows that the so-called inflationary consistency relation \cite{Liddle:1992wi} is also violated.

This paper is organized as follows. In Section (\ref{FirstSection}), we state our theory and derive the basic equations of motion. Since the unperturbed field is a constant, we consider its classical perturbations in Section (\ref{ClassicFlat}). We then quantize this theory in Section (\ref{QuantumFlat}) to show that the quantum field theory has positive definite Hamiltonian and its excitations are not tachyonic for some suitable choices of the theory parameters. In Section (\ref{FRWequations}), we put the field into a Friedman-Robertson-Walker universe, and derive the basic linear order equations for the vector field. We put these equations to work in Section (\ref{SectiondS}), solving the linear order equations in a de-Sitter universe. In Section (\ref{SectionInf}), we add an inflaton field and then compute the generated primordial perturbation spectra. Finally we conclude in Section ({\ref{Conclusion}).

\section{The Equations of Motion} \label{FirstSection}

We begin by deriving the basic equations of motion for the vector field theory. We will follow the notation of \cite{Carroll:2004ai}, where a more careful derivation is presented. In this paper we set $\hbar=c=1$; we call the gravitational constant that appear in the action $G_*$, since as we have shown in \cite{Carroll:2004ai} it differs from the observed Newton's constant $G_N$.

The action for the most general theory of a single fixed-length vector field whose derivatives are at most second order are given by Jacobson and Mattingly as \cite{Jacobson:2004ts}
\begin{equation}
S=\int dx^4 \sqrt{-g} \left(\frac{1}{16\pi G_*}R + {\cal{L}}_u \right) \label{Lagrangian}
\end{equation}
where
%\begin{equation}
%{\cal{L}}_u = K^{\mu\nu}{}_{\sigma\rho}\nabla_{\mu} u^{\sigma} \nabla_{\nu} u^{\rho}+ \lambda[u^{\mu} u_{\mu} +m^2]
%\end{equation}
\begin{equation}
{\cal{L}}_u = -\beta_1 \nabla_{\mu} u^{\nu} \nabla^{\mu} u_{\nu}-\beta_2(\nabla_{\mu} u^{\mu})^2-\beta_3 \nabla_{\mu} u^{\nu} \nabla_{\nu} u^{\mu}+ \lambda[u^{\mu} u_{\mu} +m^2].
\end{equation}
Here the $\beta_i$'s are dimensionless parameters of our theory, and $\lambda$ is a lagrange multiplier field; the vector field itself has a dimension of mass. We have dropped all terms higher than quadratic order in $u$.

Although this theory is manifestly Lorentz-invariant, its always non-vanishing vacuum violates Lorentz symmetry by picking out a dynamical frame at every point in spacetime. One can say that it has spontaneously broken Lorentz symmetry \cite{Jacobson:2000xp}; since $u^{\mu}$ is non-vanishing, there is always a dynamically preferred frame picked out by the vector field at all points in spacetime.

To see Lorentz-violating matter dynamics, in principle we need to couple this vector field to the matter action (see for example \cite{Carroll:1990vb}). In this paper we are interested in its gravitational effects as a matter field. 

The equation of motion for $\lambda$ enforces the fixed time-like norm constraint
\begin{equation}
g_{\mu\nu}u^{\mu}u^{\nu}=-m^2, \label{FixedNorm}
\end{equation}
$m$ being the squared norm of the vector field, while the equation of motion for the vector field itself is
\begin{equation}
\nabla_{\mu}J^{\mu\nu}=\lambda u^{\nu}. \label{EOMforu}
\end{equation}
where the ``current'' tensor 
\begin{equation}
J^{\mu}{}_{\nu}\equiv -\beta_1\nabla^{\mu}u_{\nu}-\beta_2\nabla_{\rho}u^{\rho}\delta^{\mu}_{\nu}-\beta_3\nabla_{\nu}u^{\mu}.
\end{equation}
By multiplying Eqn. (\ref{EOMforu}) with $u_{\nu}$ on both sides, we obtain $\lambda=-m^{-2}u_{\nu}\nabla_{\mu}J^{\mu\nu}$. Using this we find that the stress-energy tensor for this theory is \cite{Jacobson:2004ts}
\begin{eqnarray}
T_{\alpha\beta}&=&2\beta_1(\nabla_{\alpha} u^{\mu}\nabla_{\beta} u_{\mu}-\nabla^{\mu} u_{\alpha} \nabla_{\mu} u_{\beta})  \nonumber \\
&&-2[\nabla_{\mu}(u_{(\alpha}J^{\mu}{}_{\beta)})+\nabla_{\mu}(u^{\mu} J_{(\alpha\beta)})-\nabla_{\mu}(u_{(\alpha}J_{\beta)}{}^{\mu})] \nonumber \\
&&-2m^{-2}u_{\nu}\nabla_{\mu} J^{\mu\nu}u_{\alpha} u_{\beta}+g_{\alpha\beta}{\cal{L}}_u. \label{STforu}
\end{eqnarray}

\section{Classical Perturbations in Flat Space} \label{ClassicFlat}

Before we study the behaviour of the vector field in a cosmological setting, we first consider the behaviour of this vector field in Minkowski space without gravity\footnote{This limit corresponds to the setting $16\pi G_* \beta_i m^2\ll1$, see Section (\ref{ciconditions}) below.}. Since the unperturbed vector field is a constant, we investigate the evolution of its perturbations. In particular we want to put constraints on the parameters $\beta_i$, by appealing to the physical properties of its perturbations both classically and quantum mechanically. 

Our goal in this section is to find the classical solutions to the equations of motions of the vector perturbations. Along the way, we derive the Lagrangian for the vector perturbations, which we will use in Section (\ref{QuantumFlat}) when we consider the quantum version of this theory.

In Minkowski space (with metric signature $+2$) without gravity, a solution to the equations of motion (\ref{FixedNorm}) and (\ref{EOMforu}) is simply
\begin{equation}
u^{\mu}=(m,0,0,0)
\end{equation}
and
\begin{equation}
\lambda=0.
\end{equation}
To investigate its dynamics we proceed to linear order by perturbing $u$ and $\lambda$
\begin{equation}
u^{\mu}\longrightarrow \bar{u}^{\mu}+v^{\mu}~,~\lambda\longrightarrow \bar{\lambda}+\delta \lambda.
\end{equation}
We insert these expansions back into the Lagrangian Eqn. (\ref{Lagrangian}), and expand to quadratic order
\begin{equation}
{\cal{L}}_u\longrightarrow {\cal{L}}_{\bar{u}}+\delta_1 {\cal{L}} +\delta_2 {\cal{L}}
\end{equation}
where $\delta_1 {\cal{L}}$ and $\delta_2 {\cal{L}}$ contain terms of linear and quadratic order in $v$ respectively. Setting $\delta_1 {\cal{L}}=0$ will recover the equations of motion for the background variables $\bar{u}$, while variation of the quadratic order Lagrangian with respect to the perturbed variables will give us the first order perturbation equations of motion \cite{Mukhanov:1992me}.

Since we are interested in the dynamics of the perturbations, we now focus on the quadratic order Lagrangian 
\begin{equation}
\delta_2 {\cal{L}}=2  \left[-\beta_1\partial_{\mu}v^{\nu}\partial^{\mu}v_{\nu}-\beta_2(\partial_{\mu}v^{\mu})^2-\beta_3\partial^{\mu}v^{\nu}\partial_{\nu}v_{\mu}+\delta \lambda(\bar{u}^{\mu}v_{\mu}) \right]. \label{pertaction}
\end{equation}
where we have used the background result $\bar{\lambda}=0$. The equation of motion for $\delta \lambda$ is
\begin{equation}
\bar{u}^{\mu}v_{\mu}=0,
\end{equation}
and by recalling $\bar{u}^{\mu}=(m,0,0,0)$, we obtain
\begin{equation}
v^{0}=0. \label{v0iszero}
\end{equation} 
The vector field has no timelike perturbation; this follows from the fact that it is already constant in that direction by the fixed norm constraint and our choice of ansatz for its solution. 

Since $\delta \lambda$ is a lagrange multiplier, we can insert the result Eqn. (\ref{v0iszero}) back into the Lagrangian Eqn. (\ref{pertaction}) to obtain 
\begin{equation}
\delta_2 {\cal{L}}=2 \left[\beta_1\dot{v}^2-\beta_1\partial_i v^j \partial^i v_j -\beta_2 (\partial_i v^i)^2-\beta_3\partial_i v^j \partial_j v^i \right] \label{pertaction2},
\end{equation}
where the Roman indices run from 1 to 3 and overdot denotes derivative with respect to time $t$.
It is now perfectly reasonable to vary this action with respect to $v^i$ to obtain its equation of motion, since it is already in a respectable representation of the SO(3) group. However, this is a reducible representation, so under spatial rotations the components of the vector field will transform into each other. Thus it is more illuminating if we first decompose $v^{i}$ into its irreducible spin-0 and spin-1 components by
\begin{equation}
v^i\equiv \partial^{i}{\cal{V}}+N^{i}{}
\end{equation} 
where ${\cal{V}}$ is a ``scalar'' potential and $N^{i}$ is the component transverse to $\partial^i {\cal{V}}$ (i.e. $\partial_i N^i=0$). 

There are two reasons we do this. Firstly it extracts out the real dynamical degrees of freedom for the vector field, allowing us to quantize it separately. Secondly, we would like to consider its evolution in a cosmological setting, and decomposing them allows us to make contact with current literature in cosmological perturbation theory\footnote{We often use the superscripts ${}^{(0)}$, ${}^{(1)}$ and ${}^{(2)}$ to denote spin-0, spin-1 and spin-2 fields and quantities associated with them respectively. We have not seen spin-2 fields yet, but we will when we consider cosmological perturbations and gravity waves. In cosmological perturbation theory, these are often called ``scalar-type'', ``vector-type'' and ``tensor-type'' perturbations \cite{Kodama:1984bj}. To avoid phraseological disasters like ``scalar perturbations of the vector'', we do not use this terminology though we will helpfully remind the reader once in a while.}. 

With this decomposition, the Lagrangian can be written as a sum of two uncoupled pieces for the spin-0 and spin-1 components. Defining $S_i\equiv\partial_i {\cal{V}}$, it is  
\begin{equation}
\delta_2 {\cal{L}}\equiv{\cal{L}}^{(0)}+{\cal{L}}^{(1)}
\end{equation}
where
\begin{equation}
{\cal{L}}^{(0)}\equiv2 \left[\beta_1\dot{S}^2-\beta_1\partial_i S^j \partial^i S_j -\beta_2 (\partial_i S^i)^2-\beta_3 \partial_i S^j \partial_j S^i \right] \label{pertaction3spin0}
\end{equation}
is the spin-0 field Lagrangian, and
\begin{equation}
{\cal{L}}^{(1)}\equiv 2\left[\beta_1\dot{N}^2-\beta_1\partial_i N^j \partial^i N_j\right]\label{pertaction3spin1}
\end{equation}
is the spin-1 field Lagrangian. Note that we have eliminated the cross terms between $S^i$ and $N^i$, as well as the $\beta_3$ term for the spin-1 component via integration by parts. 

We now reap the first benefits of our diligence by computing the equations of motion. Varying the action with respect to $S^{i}$ and $N^i$ we get
\begin{eqnarray}
\ddot{S_i}-\frac{\beta_1+\beta_2+\beta_3}{\beta_1}\partial^2 S_i&=&0 \label{Spin0EOM}\\
\ddot{N_i}-\partial^2 N_i&=&0 \label{Spin1EOM}
\end{eqnarray}
which are simply wave equations with plane wave solutions
\begin{eqnarray}
S^i(\vec{\mathbf{k}})&\propto& e^{-ic_s^{(0)}kt+i\vec{\mathbf{k}}\cdot\vec{\mathbf{x}}} \\
N^i(\vec{\mathbf{k}})&\propto& e^{-ikt+i\vec{\mathbf{k}}\cdot\vec{\mathbf{x}}}.
\end{eqnarray}
Note that the the spin-0 component has a propagation speed of 
\begin{equation}
c_s^{(0)}{}^2=\frac{\beta_1+\beta_2+\beta_3}{\beta_1}
\end{equation}
while the spin-1 component propagates at the speed of light. If we insist that the spin-0 component does not propagate superluminally, then we obtain the following constraint:
\begin{equation}
\frac{\beta_1+\beta_2+\beta_3}{\beta_1} \leq 1 \label{c123lec1}.
\end{equation}

Since Lorentz invariance is being violated, it is not surprising that the massless spin-0 field $S_i$ may not propagate at the speed of light. Here we have imposed the condition that its propagation speed to be less than unity. Since we are already violating Lorentz symmetry, one can argue that it could propagate faster than the speed of light. However if we allow for superluminal propagation of the fields, in a curved spacetime, the theory \emph{a priori} does not exclude the possibility of the vector fields tilting in such a way that they can fit together to form a closed loop, leading to the unphysical flowing of energy around closed timelike curves.

%Note that the extra derivative couplings in the action has essentially given the spin-0 field a mass thus it does not propagate at the speed of light.
%Since Lorentz invariance is being violated, it is not surprising that the massless spin-0 field $S_i$ does not propagate at the speed of light. 

%Nevertheless, there is something not quite right in the action Eqn. (\ref{pertaction2}) : there is an overall minus sign over both the kinetic energy and the gradient energy terms. This is the first indication that the $c_i's$ has to be less than zero as we will see in the next section when we quantize the fields.

\section{Quantum Perturbations in Flat Space} \label{QuantumFlat}

A cursory look at the Lagrangian Eqn. (\ref{pertaction2}) tells us that if $\beta_1<0$, the sign of the kinetic energy is wrong which potentially will lead to quantum pathologies. In this section we will quantize the fields and show that this is indeed the case, i.e. $\beta_1>0$ if we are to have a positive definite Hamiltonian for both spin-0 and spin-1 components.

The bonus from this exercise is that we obtain a consistent way of choosing initial conditions for the vector field, a feature we will need when we put this vector field into an inflating universe in Section (\ref{SectiondS}).

\subsection{Spin-0}

We begin by quantizing the spin-0 component. From the spin-0 Lagrangian Eqn. (\ref{pertaction3spin0}), 
\begin{equation}
{\cal{L}}^{(0)}=2\left[\beta_1\dot{S}^2-\beta_1\partial_i S^j \partial^i S_j  -\beta_2 (\partial_i S^i)^2-\beta_3 \partial_i S^j \partial_j S^i \right]
\end{equation}
the canonical conjugate momentum is 
\begin{equation}
\pi_i^{(0)}\equiv \frac{\partial {\cal{L}}^{(0)}}{\partial \dot{S}^i}=4\beta_1\dot{S}_i.
\end{equation}
Promoting the variables into operators $S_i\rightarrow \hat{S}_i$, $\pi_i^{(0)}\rightarrow \hat{\pi}_i^{(0)}$, we expand them as a sum of ladder operators and their associated mode functions
\begin{eqnarray}
\hat{S}_i&=&\frac{1}{(2\pi)^{3/2}}\int d^3 k \left(\av{k}S^{(0)}_k \YLI{k}{0} + \adv{k}S^*_k{}^{(0)}\YSLI{k}{0}\right) \label{Spin0expansion}\\
\hat{\pi}^{(0)}_i&=&4\beta_1\frac{1}{(2\pi)^{3/2}}\int d^3 k \left(\av{k}\dot{S}^{(0)}_k \YLI{k}{0} + \adv{k}\dot{S}^*_k{}^{(0)}\YSLI{k}{0}\right). \label{Spin0expansion1}
\end{eqnarray}
The $\YLI{k}{0}$'s are eigenmodes of the Laplace-Beltrami operator $\partial^2$ such that $\partial^2 \YLI{k}{0}=-k^2\YLI{k}{0}$; a complete description of these eigenmodes, which are simply representations of $\mathrm{SO}(3)$, can be found in Appendix (\ref{AppA}).
%\footnote{Unfortunately, the whole eigenmode decomposition procedure is a notational nightmare and it is easy to become mired in details when reading the equations. When reading the equations, it is helpful to keep in mind that they are essentially plane waves, albeit written in a high-brow way.}. 

Following the spin-statistics theorem for integer spin fields, we impose the following equal-time commutation relations on the operators:
\begin{eqnarray}
[\hat{S}_i(\vec{\mathbf{x}}),\hat{S}^j(\vec{\mathbf{x}}')]&=&0 \\
{}[\hat{\pi}_i^{(0)}(\vec{\mathbf{x}}),\hat{\pi}^j{}^{(0)}(\vec{\mathbf{x}}')]&=&0 \\
{}[\hat{S}_i(\vec{\mathbf{x}}),\hat{\pi}^j{}^{(0)}(\vec{\mathbf{x}}')]&=&i\delta(\vec{\mathbf{x}}-\vec{\mathbf{x}}')\delta_i^j.
\end{eqnarray}
Plugging the expansion Eqn. (\ref{Spin0expansion}) into the equation of motion Eqn. (\ref{Spin0EOM}), we obtain the solution
\begin{equation}
S^{(0)}_k=A_{+}^{(0)}e^{-i c_s^{(0)}kt}+A_{-}^{(0)} e^{+i c_s^{(0)}kt} \label{Spin0Sol}
\end{equation}
where $k=\abk$.
If we choose the vacuum state $|0\rangle$ to be the one annihilated by the lowering operator $\av{k}|0\rangle=0$, then we call the $A_{+}$ term the positive frequency mode and set $A_{-}=0$. In order to normalize the ladder operator commutation relations to unity, we choose the normalization 
\begin{equation}
A_{+}^{(0)}=\frac{1}{2\sqrt{2|\beta_1|c_s^{(0)}k}}. \label{spin0vacuum}
\end{equation}
Using Eqn. (\ref{Spin0expansion}) and Eqn. (\ref{Spin0expansion1}), we find that the ladder operators obey the commutation relations
\begin{eqnarray}
[\av{k},\advp{k}]&=&  \mathrm{sgn}(\beta_1)\delta(\vec{\mathbf{k}}-\vec{\mathbf{k}}') \label{comm1} \\
{}[\av{k},\avp{k}]&=&0 \label{comm2} \\
{}[\adv{k},\advp{k}]&=& 0 \label{comm3} 
\end{eqnarray}
confirming our previous suspicion that $\beta_1>0$ for the field to have the right spin-statistics relation. In other words, we need $\beta_1>0$ for spin-0 states to have positive norm (i.e. non-ghostlike). Since $c_s^{(0)}{}^2=(\beta_1+\beta_2+\beta_3)/\beta_1\geq0$ for the field to be non-tachyonic and stable, we also find that $\beta_1+\beta_2+\beta_3\geq0$.

We can check that our choice of sign for $\beta_1$ is the right one by computing the Hamiltonian 
\begin{eqnarray}
H&=&\int d^3x \left[\hat{\pi}_i^{(0)}\dot{\hat{S}}{}^i-{\cal{L}}^{(0)} \right] \\
&=&\mathrm{sgn}(\beta_1)\frac{1}{2}\int d^3k(c_s^{(0)} k)\left[\adv{k}\av{k}+\admv{k}\amv{k}+\delta(0)\right]
\end{eqnarray}
which clearly shows that $\beta_1>0$ for it to be positive definite. 

\subsection{Spin-1}

The quantization of the spin-1 component proceeds analogously to the spin-0 component; there are just more indices to keep track of. As usual, we begin with the Lagrangian Eqn. (\ref{pertaction3spin1})
\begin{equation}
{\cal{L}}^{(1)}=2\left[\beta_1\dot{N}^2-\beta_1\partial_i N^j \partial^i N_j\right]
\end{equation}
which has the conjugate momentum
\begin{equation}
\pi_i^{(1)}\equiv \frac{\partial {\cal{L}}^{(1)}}{\partial \dot{N}^i}=4\beta_1\dot{N}_i.
\end{equation}
Promoting the variables into operators $N_i\rightarrow \hat{N}_i$, $\pi_i^{(1)}\rightarrow \hat{\pi}_i^{(1)}$, we expand them as a sum of ladder operators and their associated mode functions
\begin{eqnarray}
\hat{N}_i&=&\frac{1}{(2\pi)^{3/2}}\int d^3 k \left(\av{k}N^{(\pm 1)}_k \YLI{k}{\pm 1} + \adv{k}N^*_k{}^{(\pm 1)}\YSLI{k}{\pm 1}\right) \label{Spin1expansion}\\
\hat{\pi}^{(1)}_i&=&4\beta_1\frac{1}{(2\pi)^{3/2}}\int d^3 k \left(\av{k}\dot{N}^{(\pm 1)}_k \YLI{k}{\pm 1} + \adv{k}\dot{N}^*_k{}^{(\pm 1)}\YSLI{k}{\pm 1}\right), \label{Spin1expansion1}
\end{eqnarray}
where $\YLI{k}{\pm 1}$ are the analogous Laplace-Beltrami eigenmodes for the spin-1 field.
The ladder operators here are not the same operators from the spin-0 case; they act on a different Hilbert space. We have dropped the superscripts ${}^{(\pm 1)}$ on the ladder operators for the sake of simplicty. The right hand side is summed over both $\pm 1$ modes.
As spin-1 fields, we impose the equal time commutation relations
\begin{eqnarray}
[\hat{N}_i(\vec{\mathbf{x}}),\hat{N}^j(\vec{\mathbf{x}}')]&=&0 \\
{}[\hat{\pi}_i^{(1)}(\vec{\mathbf{x}}),\hat{\pi}^j{}^{(1)}(\vec{\mathbf{x}}')]&=&0 \\
{}[\hat{N}_i(\vec{\mathbf{x}}),\hat{\pi}^j{}^{(1)}(\vec{\mathbf{x}}')]&=&i\delta(\vec{\mathbf{x}}-\vec{\mathbf{x}}')\delta^j_i
\end{eqnarray}
on the operators.

Inserting the expansion Eqn. (\ref{Spin1expansion}) into the equation of motion Eqn. (\ref{Spin1EOM}), and choosing the vacuum such that $\av{k}|0\rangle=0$, we obtain the solution for the mode function $N_k^{(\pm 1)}$
\begin{equation}
N_k^{(\pm 1)}=\frac{1}{4\sqrt{|\beta_1|k}}{e^{-i k t}}, \label{Spin1Vacuum}
\end{equation}
noting that the propagation speed $c^{(1)}_s{}^2=1$ for the spin-1 field. Here we have dropped the negative frequency term, and again normalizing it such that the ladder operator commutation relations are normalized to unity, taking into account the fact that we sum over both $\pm 1$ modes in (\ref{Spin1expansion}) and (\ref{Spin1expansion1}). 

Using Eqn. (\ref{Spin1expansion}) and Eqn. (\ref{Spin1expansion1}), we find that the ladder operators obey
\begin{eqnarray}
[\av{k},\advp{k}]&=&  \mathrm{sgn}(\beta_1)\delta(\vec{\mathbf{k}}-\vec{\mathbf{k}}') \label{comm1a} \\
{}[\av{k},\avp{k}]&=&0 \label{comm2a} \\
{}[\adv{k},\advp{k}]&=& 0 \label{comm3a} 
\end{eqnarray}
so that, as in the spin-0 case, $\beta_1>0$ for the spin-1 field to have the right spin-statistics relation.

Finally, we can compute the Hamiltonian 
\begin{eqnarray}
H&=&\int d^3x\left[ \hat{\pi}_i^{(1)}\dot{\hat{N}}{}^i-{\cal{L}}^{(1)} \right] \\
&=&\mathrm{sgn}(\beta_1)\frac{1}{2}\int d^3k k\left[\adv{k}\av{k}+\admv{k}\amv{k}+\delta(0)\right]
\end{eqnarray}
to show that $\beta_1>0$ for it to be positive definite.

Thus for the perturbations of the LV vector field to have a consistent, non-tachyonic quantum field theory, the values of $\beta_i$ are constrained; namely $\beta_1>0$, $\beta_1+\beta_2+\beta_3\geq0$ and $(\beta_1+\beta_2+\beta_3)/\beta_1\leq1$. There is an additional constraint on $\beta_i$ that we will find when we consider the propagation of gravity waves. We will summarize all the constraints on $\beta_i$ in Section (\ref{ciconditions}) below.

\section{LV Vector Field Perturbations in an FRW Universe} \label{FRWequations}

In this section, we compute the equations of motion for the evolution of the vector field in a Friedmann-Robertson-Walker universe, adding in the effect of gravity, which we have neglected in the last two sections. The reader may want to skip this section, and refer back to it for the equations.

Since the vector field itself does not contribute any stress-energy tensor \emph{without} the presence of other matter fields in the background \cite{Carroll:2004ai}, it is inevitable that we will have to include the perturbation variables of these matter fields. 

The exception here is that of an FRW universe with the vector field and a cosmological constant. A cosmological constant contributes energy density to the stress-energy tensor in the background which the vector field will track, but it has no perturbations of its own, allowing us to solve the vector-gravity theory analytically in Section (\ref{SectiondS}).

\subsection{Background Results and Perturbation Variables}

In linear perturbation theory, the Einstein equation is split into background and perturbed components, both which are solved separately :
\begin{equation}
\bar{G}_{\mu\nu}=8\pi G_*\bar{T}_{\mu\nu}~,~\delta G_{\mu\nu}=8\pi G_*\delta T_{\mu\nu}.
\end{equation}

For simplicity, we work with a spatially flat universe; its linearly perturbed FRW metric in conformal time is
\begin{equation}
ds^2=a(\eta)^2\left[-(1+2\Phi)d\eta^2+B_i d\eta dx^i +(\gamma_{ij}+2\Psi\gamma_{ij}+2H_T{}_{ij})dx^i dx^j\right] \label{FRWperturbmetric}
\end{equation}
where $\Psi$, $\Phi$, $B_i$ and the traceless $H_T{}_{ij}$ are the four metric perturbation variables. The conformal time $\eta$ runs from $-\infty$ to $0$. $\gamma_{ij}$ is the flat spatial metric in 3 dimensions.

We have shown in a previous paper \cite{Carroll:2004ai} that the background energy density and pressure of the vector field precisely tracks those of other matter fields present in a Friedman-Robertson-Walker universe, the net effect being that the gravitational constant $G_*$ is rescaled as first noted by \cite{Mattingly:2001yd}
\begin{equation}
G_c\equiv\frac{G_*}{1+8\pi G\alpha}
\end{equation}
where we have defined 
\begin{equation}
\alpha\equiv (\beta_1+3\beta_2+\beta_3)m^2.
\end{equation}
In the same paper, we have also shown that in the Newtonian limit the vector field similarly rescales the Newton's constant by
\begin{equation}
G_N\equiv\frac{G_*}{1+8\pi G_* \delta},
\end{equation}
where
\begin{equation}
\delta\equiv -\beta_1 m^2.
\end{equation}
Note that since our experimental tests of the Newton's constant is done in the Newtonian limit, $G_N$ is the value we actually measure. As we will see in Section (\ref{ciconditions}) below, the constraints on the theory parameters $\beta_i$ mean that
\begin{eqnarray}
\alpha&\geq&0 \\
\delta&<&0.
\end{eqnarray}
In other words, the cosmological Newton's constant $G_c$ is always smaller than the measured Newton's constant $G_N$, except in the case where $m=0$.

The background Friedmann equations are 
\begin{eqnarray} 
\SH^2&=&\frac{8\pi G_c}{3}a^2 \rho_{\mathrm{m}} \label{FRWE1} \\
\adotdota-\SH^2&=&-\frac{4\pi  G_c }{3}a^2(\rho_{\mathrm{m}}+p_{\mathrm{m}}) \label{FRWE2}
\end{eqnarray}
where $\SH\equiv a'/a=aH$ while primes denote derivatives with respect to conformal time $\eta$. $\rho_{\mathrm{m}}$ and $p_{\mathrm{m}}$ are the density and pressure respectively for all other matter fields present in the universe which we model as a perfect fluid 
\begin{equation}
T^{\mathrm{m}}_{\mu\nu}=(\rho_{\mathrm{m}}+p_{\mathrm{m}})n_{\mu}n_{\nu}+p_{\mathrm{m}} g_{\mu\nu}.
\end{equation}
$n_{\mu}$ is a unit timelike vector field representing the fluid four-velocity. These matter fields can be dark matter, radiation, or some other exotic isotropic source.

Using the background result $\bar{u}^{\mu}=(m,0,0,0)$ and the fact that the fixed norm Eqn. (\ref{FixedNorm}) constrains the \emph{total} length of the vector $(\bar{u}+v)^2=-m^2$, we find that the timelike component of the vector perturbation is proportional to $\Phi$:
\begin{equation}
v^{0}=-\frac{m}{a}\Phi;
\end{equation}
i.e. as in the flat space no-gravity case in the earlier two sections, we only have three spatial dynamical degrees of freedom. For computational convenience, we write them as co-moving dimensionless variables $V^i$,
\begin{equation}
v^i\equiv\frac{m}{a}V^i. \label{SpatialComoving}
\end{equation}
We then insist that the spatial indices on $V^i$, like those on the shift perturbation $B^i$, are to be raised and lowered using the spatial metric $\gamma_{ij}$
\begin{equation}
V_i\equiv \gamma_{ij}V^j=V^i~,~B_i\equiv \gamma_{ij}B^j=B^i.
\end{equation}
Note that this definition means that the original vector perturbation obeys $v_i=ma(V_i-B_i)$. 

\subsection{The Bardeen Decomposition}

In an FRW background which is spatially homogenous and isotropic, the perturbation variables can be decomposed into spin-0, spin-1 and spin-2 components \cite{Bardeen:1980kt}. If further, the background matter fields themselves do not break spatial SO(3) symmetry, the great advantage of this decomposition is that each of the components are decoupled from one another, thus allowing us to solve for them separately. Fortunately, this is indeed the case in our theory, since the background vector field $\bar{u}^{\mu}$ only has timelike components (unlike the case in \cite{Armendariz-Picon:2004pm}, for example).

In this section, we will write down the equations for each component. The equations here may seem daunting, but they are straightforward to derive: just insert the perturbed metric (\ref{FRWperturbmetric}) and vector field $u^{\mu}$ into equations (\ref{EOMforu}) and (\ref{STforu}) from Section (\ref{FirstSection}) and then keep all terms to linear order. The mode equations for the Einstein tensor can be found in the literature \cite{Kodama:1984bj} so we will not list them here. We will put these equations to work only in the following sections, so one can skip this section and just refer back to it for the equations.

Using the eigenmodes described in Appendix (\ref{AppA}), we can expand the variables as follows
\begin{eqnarray}
\Phi&=&\sum_k \Phi Y^{(0)} \\
\Psi&=&\sum_k \Psi Y^{(0)} \\
V^i&=&\sum_k \sum_{m=0,1} V^{(\pm m)} Y^i{}^{(\pm m)}\\
B^i&=&\sum_k \sum_{m=0,1} B^{(\pm m)} Y^i{}^{(\pm m)} \\
H_T^{ij}&=&\sum_k \sum_{m=0,1,2} H_T^{(\pm m)} Y^{ij}{}^{(\pm m)}
\end{eqnarray}
where we have dropped the implicit subscript $k$ on the right hand side.
Plugging these back into the equation of motion Eqn. (\ref{EOMforu}) and the stress-energy tensor Eqn. (\ref{STforu}), we obtain their corresponding spin-0, spin-1 and spin-2 components. The equations in real space are listed in Appendix (\ref{AppC}).

\subsubsection{Spin-0 ``Scalar-type'' perturbations} \label{spin0tensor}

The spin-0 component of the equation of motion (\ref{EOMforu}) is
\begin{eqnarray}
&&\left\{\right. -6\frac{\alpha}{m^2}\SH^2\Phi+6\beta_2\left[\adotdota+\SH^2\right]\Phi+a^2\delta \lambda+\SH\left[(2\beta_1+\beta_2+2\beta_3)(kV^{(0)}+3\Psi')\right. \nonumber\\
&&-\left.\beta_3(kV^{(0)}-kB^{(0)})+3\beta_2\Phi'\right]+\beta_3(k^2\Phi+kB'^{(0)}-kV'^{(0)})\nonumber \\
&&-\left.\beta_2(kV'{}^{(0)}+3\Psi'')\right\}Y^{(0)}=0
\end{eqnarray}
for $\nu=0$, and
\begin{eqnarray}
&&\left\{\left[-2\frac{\alpha}{m^2}\SH^2+\frac{\alpha}{m^2}\adotdota+\beta_1\adotdota\right](B^{(0)}-V^{(0)})-\SH\left[\left(\beta_1+\frac{\alpha}{m^2}\right)k\Phi\right. \right. \nonumber \\
&&-\left. 2\beta_1(-B'{}^{(0)}+V'{}^{(0)})\right]+(\beta_1+\beta_2+\beta_3)k^2V^{(0)}-\frac{2}{3}(\beta_1+\beta_3)kH'_T{}^{(0)}-\beta_1k\Phi' \nonumber \\
&&+\left.\frac{\alpha}{m^2}k\Psi'-\beta_1(B''{}^{(0)}-V''{}^{(0)})\right\}Y_i^{(0)}=0 \label{spin0spaceEOM}
\end{eqnarray}
for $\nu=i$.

The components of the stress-energy tensor for the spin-0 fields are
\begin{eqnarray}
\delta T{}^0{}_0^{(0)}&=&2\left(\frac{m}{a}\right)^2\left\{-3\frac{\alpha}{m^2}\right.\SH^2\Phi+\beta_1a^{-1}\left[ak(B^{(0)}-V^{(0)})\right]' \nonumber \\
&&+\left. \beta_1k^2\Phi+\frac{\alpha}{m^2}(kV^{(0)}+3\Psi')\SH\right\}Y^{(0)} \label{T000}\\
\delta T^0{}_i^{(0)}&=&2\left(\frac{m}{a}\right)^2\left\{ \left[-2\frac{\alpha}{m^2}\SH^2+\frac{\alpha}{m^2}\adotdota-\beta_1\adotdota\right](V^{(0)}-B^{(0)})\right. \nonumber \\
&& +\left. \beta_1 k \frac{(a\Phi)'}{a}-\beta_1a^{-2}\left[a^2(V'{}^{(0)}-B'{}^{(0)})\right]\right\}Y_i^{(0)} \label{T00i}\\
\delta T^i{}_j^{(0)}&=&2\left(\frac{m}{a}\right)^2\left\{\frac{\alpha}{m^2}\left[\SH^2-2\adotdota\right]\Phi-\SH\frac{\alpha}{m^2}\Phi'\right. \nonumber \\
&&+\left. a^{-2}\left[a^2(\beta_2kV^{(0)}+\frac{\alpha}{m^2}\Psi'+(\beta_1+\beta_3)\frac{k}{3}V^{(0)})\right]'\right\}Y^{(0)}\delta^i{}_j \nonumber\\
&&-2\left(\frac{m}{a}\right)^2\left\{(\beta_1+\beta_3)a^{-2}\left[a^2(kV^{(0)}-H'_T{}^{(0)})\right]'\right\}Y^i{}_j{}^{(0)}. \label{T0ij}
\end{eqnarray}

\subsubsection{Spin-1 ``Vector-type'' perturbations}

The spin-1 component of the equation of motion Eqn. (\ref{EOMforu}) is
\begin{eqnarray}
&&\left\{\left[-2\frac{\alpha}{m^2}\SH^2+\frac{\alpha}{m^2}\adotdota-\beta_1\adotdota\right](B^{(\pm1)}-V^{(\pm1)})\right. \nonumber \\
&&+2\beta_1\SH(V'{}^{(\pm1)}-B'{}^{(\pm 1)})+\frac{1}{2}(\beta_3-\beta_1)k^2B^{(\pm 1)}+\beta_1k^2V^{(\pm 1)} \nonumber \\
&& \left. -(\beta_1+\beta_3)\frac{k}{2}H'_T{}^{(\pm 1)}-\beta_1(B''{}^{(\pm 1)}-V''{}^{(\pm 1)})\right\}Y_i^{(\pm 1)}=0.
\end{eqnarray}

The components of the stress-energy tensor for the spin-1 fields are
\begin{eqnarray}
\delta T^0{}_0^{(\pm 1)}&=&0 \label{T100} \\
\delta T^0{}_i^{(\pm 1)}&=&2\left(\frac{m}{a}\right)^2\left\{\left[-2\frac{\alpha}{m^2}\SH^2+\frac{\alpha}{m^2}\adotdota-\beta_1\adotdota\right](V^{(\pm 1)}-B^{(\pm 1)})\right. \nonumber \\
&& -\beta_1a^{-2}\left[a^2(V'{}^{(\pm 1)}-B'{}^{(\pm 1)})\right]' \nonumber \\
&&\left. +\frac{1}{2}(\beta_1-\beta_3)k^2(B^{(\pm 1)}-V^{(\pm 1)})\right\}Y_i^{(\pm 1)} \label{T10i} \\
\delta T^i{}_j^{(\pm 1)}&=&2\left(\frac{m}{a}\right)^2\left\{a^{-2}\left[a^2(\beta_1+\beta_3)(-kV^{(\pm 1)}+H'{}_T^{(\pm 1)})\right]'\right\}Y^i{}_j{}^{(\pm 1)}. \label{T1ij}
\end{eqnarray}

\subsubsection{Spin-2 ``Tensor-type'' perturbations}

There is no spin-2 equation of motion for the vector field, since it has at most spin-1 components. However, its stress-energy is non-zero since it contains curvature terms :
\begin{eqnarray}
\delta T^0{}_0^{(\pm 2)}&=&\delta T^0{}_i^{(\pm 2)}=0 \\
\delta T^i{}_j^{(\pm 2)}&=&2\left(\frac{m}{a}\right)^2\left\{a^{-2}\left[a^2(\beta_1+\beta_3)H'{}_T^{(\pm 2)}\right]'\right\}Y^i{}_j{}^{(\pm 2)}. \label{T2ij}
\end{eqnarray}

\section{LV Vector Field Perturbations in de-Sitter Space} \label{SectiondS}

The paradigmatic theory of early universe is currently the theory of inflation \cite{Guth:1981zm}, which postulates that the early universe undergoes an accelerated phase of expansion. One key to its success is that it provides a mechanism for consistently choosing initial conditions to produce a scale-invariant spectrum of cosmological perturbations that eventually will grow into galaxies and things like human beings \cite{Bardeen:1983qw}. 

This mechanism is based on the idea that quantum fluctuations in a near de-Sitter background can be ``blown up'' in their amplitude such that they become classical perturbations after they cross the Hubble radius. Thus, the initial conditions for the perturbations are quantum mechanical in nature; one chooses a suitable vacuum for the initially quantum field when the physical length of the mode is very small, and then follows the evolution of this mode until it freezes out. The classical amplitude of this mode is then given to us by the 2-pt correlation function of the quantum field \cite{Mukhanov:1992me}.

In general, inflation is driven by some dynamical field called the inflaton; in the usual case it is a scalar field. The inflaton itself has perturbations, so a coupled vector-inflaton-gravity theory is not easily solved. Before we do that in the next section, as a warm up we tackle the simpler case of inflation driven by a cosmological constant, which has no perturbations. In this case, the vector-gravity perturbation equations are analytically solvable in closed form. 

The solutions to the various equations are straightforward, though it takes some algebraic manipulation to write them in a useful form; here we sketch the derivation. One can of course skip them and just look right ahead at the solution in Eqns. (\ref{dSSpin2Sol}), (\ref{dSSpin1Sol}) and (\ref{dSSpin0Sol}). We will turn the order around this time and compute the solutions for spin-2 perturbations first, before finishing with the spin-1 and spin-0 perturbation solutions.

\subsection{Evolution of spin-2 perturbations in de-Sitter space}

Although the LV vector field has no spin-2 component, as we have shown in Section (\ref{FRWequations}) it has a non-zero stress-energy tensor; its very presence will modify the evolution of gravity waves.

Spin-2 equations are gauge invariant\footnote{Since gauge transformations are generated by diffeomorphisms which in turn are generated by a vector field, this means that there are no spin-2 gauge transformations.}. By using the $i-j$ component of the Einstein equation, we obtain the equation of motion for the spin-2 gravity wave modes
\begin{equation}
H''_T{}^{(\pm 2)}+2\SH H'_T{}^{(\pm 2)}+\frac{k^2}{1-\gamma}H_T^{(\pm 2)}=0 \label{dSSpin2Sol} 
\end{equation}
\begin{equation}
\gamma \equiv 16\pi G_* m^2 (\beta_1+\beta_3) 
\end{equation}
which is the same equation for propagation of gravity waves in de-Sitter space without the vector field except that the propagation speed has been rescaled 
\begin{equation}
c_s^{(2)}{}^2=\frac{1}{1-\gamma}.
\end{equation}
This gives us an extra constraint that $\beta_1+\beta_3\leq0$ if they are not to propagate faster than the speed of light. We can understand this by realizing the the vector field has changed the permeability of ``empty space''. This was first noted by the authors of \cite{Jacobson:2004ts} in a Minkowski space setting, whose results for the propagation speeds we also recover for spin-0 and spin-1 fields in the following sections (by setting $a=\mathrm{constant}$). Note that their ``small $c_i$'' limit corresponds to $16\pi G_* \beta_i m^2\ll1$ here.
% \footnote{Though we hasten to add that there is nothing unphysical about faster than light propagating of gravity waves : we have not detected any.}

Since gravity wave perturbations freeze out at the sound horizon $c_s^{(2)} k \eta=1$, they will freeze out earlier compared to the usual case of $k\eta=1$, leading to a modification of the amplitude of the spectrum. We leave the computation of this spectrum to Section (\ref{SectionInf}).

\subsection{Evolution of spin-1 perturbations in de-Sitter space}

Unlike the spin-2 equations, the spin-1 Einstein equation has two gauge degrees of freedom, and one can use this freedom to simplify the equations somewhat. Following \cite{Hu:1997hp}, we choose
\begin{equation}
H_T^{(\pm 1)}=0
\end{equation}
i.e. we consider the perturbations on hypersurfaces that are shear-free.

The spin-1 $i-j$ component of the Einstein equation gives us the constraint between the shift $B^{(1)}$ and the spin-1 perturbation $V^{(1)}$,
\begin{equation}
B^{(\pm 1)}=\gamma V^{(\pm 1)}.
\end{equation}
Inserting this into $0-i$ component of the Einstein equation results in 
\begin{eqnarray}
&&\left[2\frac{\alpha}{m^2}\SH^2-\frac{\alpha}{m^2}\adotdota+\beta_1\adotdota\right]V^{(\pm 1)}+ \nonumber\\
&&+\frac{1}{2}\left[(\beta_1-\beta_3)+\frac{\beta_1+\beta_3}{1-\gamma}\right]k^2 V^{(\pm 1)}+\beta_1(2\SH V^{(\pm 1)}{}'+V^{(\pm 1)}{}'')=0, 
\end{eqnarray}
which after defining 
\begin{equation}
\xi^{(\pm 1)}\equiv aV^{(\pm 1)},
\end{equation}
becomes
\begin{equation}
\xi^{(\pm 1)}{}''+\frac{1}{2}\left[(1-\beta_3/\beta_1)+\frac{1+\beta_3/\beta_1}{1-\gamma}\right]k^2\xi^{(\pm 1)}+A\frac{\alpha}{m^2\beta_1}\xi^{(\pm 1)}=0. \label{dSSpin1Sol}
\end{equation}
Here we have defined the useful background quantity 
\begin{equation}
A\equiv\left[2\SH^2-\adotdota\right]=4\pi G_c\left[a^2\frac{\rho_{\mathrm{m}}+p_{\mathrm{m}}}{2}\right]
\end{equation}
using the Friedman equations (\ref{FRWE1}) and (\ref{FRWE2}).

In de-Sitter space the pressure of the cosmological constant is equal to minus the energy density, $p_{\mathrm{m}}=-\rho_{\mathrm{m}}$, so that $A=0$; thus the last term in Eqn. (\ref{dSSpin1Sol}) vanishes leaving us with an oscillatory solution for $\xi^{(\pm 1)}$. This means that
\begin{equation}
V^{(\pm 1)}=\frac{a(\eta_0)}{a}\frac{1}{4m\sqrt{c^{(\pm 1)}_s |\beta_1|k}}e^{i c_s^{(\pm 1)}k \eta}
\end{equation}
with the propagation speed
\begin{equation}
 c_s^{(\pm 1)}{}^2\equiv\frac{1}{2}\left[(1-\beta_3/\beta_1)+\frac{1+\beta_3/\beta_1}{1-\gamma}\right].
\end{equation}
The normalization is set by choosing an appropriate early time $\eta_0$ for the mode's ``birth'', and then matching $V^{(\pm 1)}$ to the quantum vacuum state $N_k^{(\pm 1)}$ we previously found in Eqn. (\ref{Spin1Vacuum}) of Section (\ref{QuantumFlat}). This matching is reasonable because at very short wavelengths where presumably when these modes are ``created'', they do not see the large-scale curvature of space time. In this case, we can make this matching when $(a(\eta_0)/k)/H^{-1}\ll 1$ where $H^{-1}$ is the de-Sitter radius (or the scale of inflation in a near de-Sitter cosmology). The vector norm $m$ is there because we have factored it out in our change of variables in Eqn. (\ref{SpatialComoving}). 

Notice that the propagation speed here is not unity as in the flat space case: the presence of gravity has changed the value somewhat (although the condition $\beta_1+\beta_3\leq0$ implied that $c_s^{(\pm 1)}{}^2\leq1$). This is due to the fact that the vector field is no longer propagating through ``vacuum'', but instead has to traverse through a ``medium'' of backreacting metric perturbations; one can check this by turning off gravity (setting $G_*=0$) and recovering $c_s^2{}^{(\pm 1)}=1$. (See also \cite{Jacobson:2004ts}.)

The spin-1 component of the shift perturbation $B^{(\pm 1)}$ sources both E-type and B-type polarization anisotropies of the CMB, the latter dominating the former by a factor of 6 \cite{Hu:1997hp}. Thus the quantity of interest that we want to compute is the spectrum of spin-1 perturbations, given by the two point correlation function
\begin{equation}
\langle 0|B^{(\pm 1)}(\vec{x})B^{(\pm 1)}(\vec{x}+\vec{r}) |0\rangle = \gamma^2\int \frac{dk}{k}\frac{\sin (kr)}{kr}\frac{1}{2\pi^2}k^3|V^{(\pm 1)}|^2.
\end{equation}
But since $V^{(\pm 1)}\propto a^{-1}$ and $a$ grows exponentially in de-Sitter space, $V^{(\pm 1)}$ can only decay. Unlike the inflaton, the vector field does not freeze out, leaving us with nothing but a bunch of complicated equations. In addition, if the inflaton field is a scalar field which contributes no spin-1 perturbations, this result is general: \emph{scalar field driven inflation with a fixed-norm LV vector field does not generate a spin-1 spectrum of perturbations.}

In fact after the end of inflation, the first term in Eqn. (\ref{dSSpin1Sol}) will decay as $a^{-1}$ and $a^{-2}$ in matter and radiation dominated eras respectively, leaving our oscillatory solution for $\xi^{(\pm 1)}$ unchanged thus even if these modes are created during the CMB era they will decay away \footnote{This is technically only true for the usual CMB fluids such as photons, baryons and dark matter. If we have \emph{other} vector sources, the evolution of $V^{(\pm 1)}$ will be non-trivial.}. 

Thus we have shown that the vector field cannot source spin-1 type perturbations, and therefore will not generate B-type polarization modes in the CMB. 

\subsection{Evolution of spin-0 perturbations in de-Sitter space} \label{sectionspin0dS}

Analogous to the spin-1 equations, the spin-0 equations have two gauge degrees of freedom, which we fix by choosing
\begin{equation}
B^{(0)}=0~,~H_T^{(0)}=0
\end{equation}
which is sometimes called the newtonian gauge or longitudinal gauge.
%\footnote{Readers who are using the Einstein equations from \cite{Kodama:1984bj} should be warned that there the authors have chosen $H_T^{(0)}{}'=0$ instead in their choice for the longitudinal gauge, leaving them with a residual degree of freedom to choose any value for $H_T^{(0)}=\mathrm{constant}$}. 

Using the off-diagonal term of the $i-j$ Einstein equation, we obtain the constraint
\begin{equation}
k^2(\Phi+\Psi)=\gamma a^{-2}(a^2 k V^{(0)})',\label{spin0constraint}
\end{equation}
which we can then plug into the $0-i$ component to obtain
\begin{eqnarray}
a^{-1}(ak\Phi)'&=&\frac{1}{2-16\pi G_* m^2 \beta_1}\left\{(-16\pi G_* m^2\beta_1+2\gamma)\frac{\xi^{(0)}{}''}{a} \right. \nonumber \\
&&\left. -(2\gamma+16\pi G_*\alpha)\left[2\SH^2-\adotdota\right]\frac{\xi^{(0)}}{a}\right\}
\end{eqnarray}
where we have similarly defined
\begin{equation}
\xi^{(0)}\equiv aV^{(0)}.
\end{equation}
Inserting this result into the space component of the equation of motion Eqn. (\ref{spin0spaceEOM}), together with the constraint Eqn. (\ref{spin0constraint}), we obtain
the following wave equation
\begin{equation}
\xi^{(0)}{}''+\frac{(\beta_1+\beta_2+\beta_3)}{Z_2}k^2\xi^{(0)}+\frac{Z_1}{Z_2}A\xi^{(0)}=0 \label{dSSpin0Sol},
\end{equation}
where
\begin{eqnarray}
Z_1&\equiv& \frac{\alpha/m^2-\beta_1\gamma}{ G_c /G_* -8\pi  G_c  m^2 \beta_1}\\
Z_2&\equiv&\frac{\beta_1(1-\gamma)}{ G_c /G_* -8\pi  G_c  m^2 \beta_1}.
\end{eqnarray}
Comparing (\ref{dSSpin0Sol}) to the spin-1 solution Eqn. (\ref{dSSpin1Sol}), we see that except for a different propagation speed
\begin{equation}
c^{(0)}_s{}^2\equiv \frac{(\beta_1+\beta_2+\beta_3)}{Z_2}
\end{equation}
and the coefficient in front of the last term on the left hand side, they are identical. One can check that the constraints on $\beta_i$ (to be listed below) means that the propagation speed $c^{(0)}_s\leq 1$ and that it reduces to the flat space case $(\beta_1+\beta_2+\beta_3)/\beta_1$ when gravity is turned off by setting $G_*=0$. 

In de-Sitter space $p_{\mathrm{m}}=-\rho_{\mathrm{m}}$ so that $A=0$, the last term in Eqn. (\ref{dSSpin0Sol}) vanishes and we have the solution 
\begin{equation}
V^{(0)}=\frac{a(\eta_0)}{a}\frac{1}{2m\sqrt{2c_s|\beta_1|k}}e^{i c_s k\eta}, \label{dSSolSpin0}
\end{equation}
where again the normalization is chosen to match the quantum vacuum state $S_k$ (Eqn. (\ref{spin0vacuum})) at some very early ``birth'' time $\eta_0$. 

We have shown that the spin-0 perturbation will decay exponentially in an inflationary phase and contribute no spectrum of density perturbations on the sky. However, in the radiation and matter dominated eras, the spin-0 equation of motion no longer behaves like Eqn. (\ref{dSSpin0Sol}) since it will couple via gravity to spin-0 perturbations of the other fluids, leading to complicated coupled equations which cannot be solved analytically in general. 

\subsection{What are the allowed values for $\beta_i$?} \label{ciconditions}

Before we proceed to the next Section, let us summarize our findings on the constraints on the theory parameters $\beta_i$:
\begin{enumerate}
\item Subluminal propagation of spin-0 field : $(\beta_1+\beta_2+\beta_3)/\beta_1\leq1$. \label{cond1}
\item Positivity of Hamiltonian : $\beta_1>0$. \label{cond2}
\item Non-tachyonic propagation of spin-0 field : $(\beta_1+\beta_2+\beta_3)/\beta_1\geq0$. \label{cond3}
\item Subluminal propagation of spin-2 field : $\beta_1+\beta_3\leq0$. \label{cond4}
\end{enumerate}

Conditions \ref{cond2} and \ref{cond3} imply that $\beta_1+\beta_2+\beta_3\geq0$, conditions \ref{cond1} and \ref{cond2} imply that $\beta_2+\beta_3\leq0$, while conditions \ref{cond2},~\ref{cond3} and \ref{cond4} imply that $\beta_2\geq0$. These constraints form an infinite length wedge in the parameter space spanned by $(\beta_1,\beta_2,\beta_3)$ as shown in Fig. (\ref{fig1}).

\begin{figure}[ptbh]
\myfigure{4in}{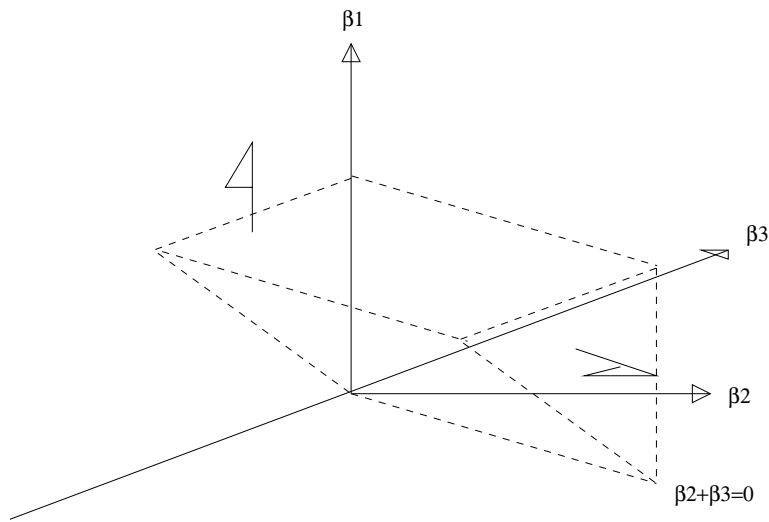}
\caption{Figure showing the allowed values for $\beta_i$. The arrows indicate extension to infinity. The slope of the wedge makes a 45 degrees angle with the $\beta_2-\beta_3$ plane. An example of a valid choice of parameters is $\beta_1=1$, $\beta_2=2.5$, $\beta_3=-3$.}
\label{fig1}
\end{figure}

The above conditions imply that the background energy density of the vector field in an FRW background 
\begin{equation}
\rho_{u}=-3\alpha H^2=-3(\beta_1+3\beta_2+\beta_3)m^2H^2\leq0
\end{equation}
is non-positive since $\alpha\geq0$. As discussed in the a companion paper \cite{Carroll:2004ai}, this is not a disaster since the background vector has no dynamical degrees of freedom. This is akin to the negative energy density of a negative cosmological constant. As shown in the previous section, for suitable choices of the parameters $\beta_i$, the quantum field theory of the propagating perturbative degrees of freedom is perfectly consistent. 

We note here that the conditions above are valid in the regime where $16\pi G_* \beta_i m^2 \ll 1$; we can understand this by noting that in this regime the energy density of vector field becomes small and thus the geometry of spacetime approaches that of Minkowski where we have derived our constraints. While we have shown above that the various propagation speeds are still subluminal in a de-Sitter background, there is no guarantee that this would be true in general. Indeed outside this regime, the modes may become tachyonic\footnote{We thank Ted Jacobson for pointing this out to us.}

\section{LV Vector Fields in Inflation} \label{SectionInf}

One can try to understand the previous section by analogy to the exact tracker solution of the background vector field: the perturbations need to track something. If the background has no matter fields and thus is Minkowski, the stress-energy of the vector field vanishes identically. Similarly, in the absence of other matter perturbations; the vector field perturbations follow the behaviour of metric perturbations, which in de-Sitter spacetime have been shown to exponentially decay \cite{Hawking:1983my}. 

It then follows that if we drive inflation with an inflaton instead of just a plain cosmological constant, the vector field might track the evolution of the inflaton perturbations, preventing it from decaying away. In this section, we investigate this possibility.

The simplest models of inflation are those driven by a slowly-rolling scalar field (see for example \cite{Kolb:1990vq}). Unfortunately, unlike the previous section, the equations become intractable to solve in exact closed form. Nevertheless, in Section (\ref{approximate}), we derive approximate solutions for the evolution of the metric perturbation $\Phi$. We find that, like its background counterpart, the inflaton perturbation acts as a source for the vector field, preventing it from decaying away.

We end our investigation into observability of LV vector fields by computing both primordial spectra for the density perturbation and the gravitational wave spectrum in Section ({\ref{spin2spectrum}), showing that both spectra are modified from the usual values in potentially detectable ways.

\subsection{Approximate solution to the Metric Perturbation $\Phi$} \label{approximate}

The observed spectrum of temperature anisotropies in the CMB \cite{Spergel:2003cb} is given by the two point correlation function of the density perturbation $\delta \rho/\rho$ at the time of recombination. The density perturbation itself is sourced by a spectrum of metric perturbation $\Phi$ modes, which remain constant until they enter the Hubble radius, at which point they oscillate to generate the observed features.

Thus it is of interest to us to investigate the evolution and the eventual large scale solution to $\Phi$. In this section, we derive the approximate solutions to $\Phi$, and then in the next section we use these solutions to calculate the spectrum of perturbations generated by inflation with the presence of the vector field.

We note in passing that since the inflaton field is a scalar, it sources no spin-1 type perturbations and thus the spin-1 perturbation evolves exactly as in the previous section. Since we are considering only spin-0 quantities here, we drop the ${}^{(0)}$ superscript. 

Consider a scalar field theory acting as the inflaton 
\begin{equation}
S_{\phi}=\int d^4x \sqrt{g} \left(-\frac{1}{2}(\partial\phi)^2-\tilde{V}(\phi)\right).
\end{equation}
The exact form of $\tilde{V}(\phi)$ is not important in what we are considering; suffice to say that it is almost flat (see for example \cite{Liddle:2000cg}). For this action, the scalar field equation of motion is 
\begin{equation}
\phi''+2\SH\phi'+a^2\frac{d \tilde{V}}{d\phi}=0, \label{scalarbackgroundEOM}
\end{equation}
where the scalar is assumed to be spatially homogenous as required by the symmetry of the background FRW metric.

The background stress-energy tensor of this scalar field has the following components
\begin{eqnarray}
T^0_{\phi}{}_0&=&-\rho_{\phi}=-\frac{1}{2}\phi'{}^2a^{-2}-\tilde{V}\\
T^i_{\phi}{}_j&=&p_{\phi}\delta^i_j=\left(\frac{1}{2}\phi{}'^2a^{-2}-\tilde{V}\right)\delta^i_j.
\end{eqnarray}
Thus if the potential $\tilde{V}$ is very flat, then the scalar field is slowly-rolling $\phi^2{}'\ll \tilde{V}a^2$ and thus $p_{\phi}\approx -\rho_{\phi}$ leading to an accelerating phase of cosmic expansion.

Perturbing this field $\phi\longrightarrow \bar{\phi}+\delta \phi$, we find its linear order spin-0 stress-energy tensor is
\begin{eqnarray}
\delta T^0_{\phi}{}_0&=&\frac{1}{a^2}\left[\bar{\phi}'^2\Phi-\bar{\phi}'\delta \phi'-\frac{\partial \tilde{V}}{\partial \bar{\phi}}a^2\delta \phi \right]Y \\
\delta T^0_{\phi}{}_i&=&\frac{1}{a^2}\bar{\phi}'k\delta \phi Y_i \\
\delta T^i_{\phi}{}_j&=&-\frac{1}{a^2}\left[\bar{\phi}'^2\Phi-\bar{\phi}'\delta \phi'+\frac{\partial \tilde{V}}{\partial \bar{\phi}}a^2\delta  \phi \right]Y\delta^i{}_j.
\end{eqnarray}

Combining these with the vector stress-energy tensor components Eqns. (\ref{T000}), (\ref{T00i}), (\ref{T0ij}) from Section (\ref{spin0tensor}) and the Einstein tensor, we obtain 
\begin{eqnarray}
4\pi G_c \left(-\bar{\phi}'\delta \phi'-\frac{\partial\tilde{V}}{\partial\bar{\phi}}a^2\delta \phi \right)&=&\left[\SH^2+\adotdota\right]\Phi-3\SH\Psi'-\frac{ G_c }{G_*}k^2\Psi-8\pi G_c \beta_1 m^2 k^2\Phi\nonumber \\
&&+8\pi G_c \beta_1m^2\left(kV'+k\SH V\right)-\alpha\SH kV \label{totalEE00} \\
\frac{1}{8\pi G_*}\left(k\SH\Phi-k\Psi'\right)&=&\frac{k}{2}\bar{\phi}'\delta \phi-\alpha A V+\beta_1m^2a^{-1}(ak\Phi)'-\beta_1m^2\frac{\xi''}{a} \label{totalEE0i} \\
4\pi G_c \left(\bar{\phi}'\delta \phi'-\frac{\partial\tilde{V}}{\partial \bar{\phi}}a^2\delta \phi \right)&=&\left[\SH^2+\adotdota\right]\Phi+\SH\Phi'-2\SH\Psi'-\Psi'' \nonumber \\
&&-\frac{8\pi G_c m^2}{\gamma}(\beta_1+\beta_2+\beta_3)k^2(\Phi+\Psi) \label{totalEEij}
\end{eqnarray}
which are the $0-0$, $0-i$ and $i-i$ components of the Einstein equation respectively.

Our goal is to write down the two evolution equations for the coupled $\Phi$ and $\xi=aV$ fields, by eliminating $\Psi$ and $\delta \phi$. We first replace every instance of $\Psi$ with $\Phi$ and $\xi$ by using the constraint equation (\ref{spin0constraint}). The first equation is then simply the equation of motion (\ref{spin0spaceEOM})
\begin{equation}
\xi''+\frac{(\beta_1+\beta_2+\beta_3)m^2}{\beta_1m^2+\gamma\alpha}k^2\xi+\frac{\alpha(1-\gamma)}{\beta_1 m^2+\gamma\alpha}A\xi=\frac{\beta_1m^2+\alpha}{\beta_1m^2+\gamma\alpha}k(a\Phi)', \label{perteqn1}
\end{equation}
where now the background quantity $A$ is given by
\begin{equation}
A\equiv\left[2\SH^2-\adotdota\right]=4\pi  G_c \phi'^2=4\pi G_c\left[a^2\frac{\rho_{\phi}+p_{\phi}}{2}\right].
\end{equation}

Next we obtain the second evolution equation by first adding Eqn. (\ref{totalEE00}) and Eqn. (\ref{totalEEij}) to eliminate $\delta \phi'$. To write $\delta \phi$ in terms of the $\Phi$ and $\xi$ variables, we use the equation of motion Eqn. (\ref{spin0spaceEOM}), combined with Eqn. (\ref{totalEE0i}), and the background equation (\ref{scalarbackgroundEOM}), to obtain the following relation
\begin{eqnarray}
-8\pi  G_c \frac{d \tilde{V}}{d\phi}a^2\delta \phi &=&\left(4\SH+\frac{A'}{A}\right)\left[\left(\SH \Phi+\Phi'\right)-\frac{\gamma}{k}\frac{\xi''}{a}\right. \nonumber \\
&&\left.-\left(8\pi  G_c (\beta_1+\beta_2+\beta_3)m^2k - \frac{\gamma}{k}A \right)\frac{\xi}{a}\right] \label{dVdphi}.
\end{eqnarray}
Putting everything together, and eliminating the $d\tilde{V}/d\phi$ term in Eqn. (\ref{dVdphi}) using Eqn. (\ref{scalarbackgroundEOM}), we get the second evolution equation
\begin{eqnarray}
&&\Phi''+\left(2\SH-\frac{A'}{A}\right)\Phi'+\left[-2\SH^2+2\adotdota-a\frac{A'}{A}\SH\right]\Phi+\left(\frac{ G_c }{G_*}-8\pi  G_c m^2 \beta_1\right)k^2\Phi \nonumber \\
&&+8\pi  G_c \left[-(\beta_1+\beta_2+\beta_3)m^2+\beta_1m^2-\frac{\gamma}{8\pi G_*}\right]k\frac{\xi'}{a}+8\pi G_c m^2(\beta_1+\beta_2+\beta_3)\frac{A'}{A}k\frac{\xi}{a} \nonumber \\
&&+\frac{\gamma}{k}\left(2\frac{A'}{A}\frac{\xi''}{a}-\frac{\xi'''}{a}+A\frac{\xi'}{a}\right)=0. \label{perteqn2}
\end{eqnarray}

To further simplify our analysis below, we set $A\longrightarrow 0$, i.e. the background is very near to de-Sitter space. This means that $a''/a\sim 1/\eta^2$, and $a'/a\sim A'/A\sim 1/\eta$, with $\sim$ meaning ``scales as''.

\subsubsection{Short wavelength solution}

Consider the evolution of the perturbations at short wavelengths $k\eta\gg1$. Unlike the de-Sitter case, $V=\xi/a$ no longer decays away since the presence of the inflaton perturbation acts as a source. We can see that by using Eqn. (\ref{totalEE0i}) and doing similar computations to Section (\ref{sectionspin0dS}) to obtain 
\begin{equation}
\frac{\xi''}{a}+\frac{(\beta_1+\beta_2+\beta_3)}{Z_2}k^2\frac{\xi}{a}=\frac{\beta_1 m^2+\alpha}{\beta_1 m^2(1-\gamma)}4\pi G_c k \bar{\phi}'\delta \phi. \label{xiwithphi}
\end{equation}
The source term on the right hand side is proportional to $k$, and thus the inflaton perturbation only sources the vector perturbation at small scales. In contrast, the de-Sitter background case Eqn. (\ref{dSSpin0Sol}) is always sourceless.

To investigate the evolution of $\Phi$ however, it is more convenient to use Eqn. (\ref{perteqn1}) which we can solve (recall that we have set $A=0$)
\begin{equation}
\xi=\left[D_1 \cos(\tilde{c_s}k\eta)+D_2\sin(\tilde{c_s}k\eta)\right]+\frac{\beta_1m^2+\alpha}{\beta_1m^2+\alpha\gamma}\int k(a\Phi)d\eta, \label{xisolution}
\end{equation}
where 
\begin{equation}
\tilde{c_s}^2\equiv \frac{(\beta_1+\beta_2+\beta_3)m^2}{\beta_1m^2+\alpha\gamma}
\end{equation}
and $D_1$ and $D_2$ are integration constants. The bracketed first term on the right hand side is just the usual oscillating solution (albeit with a modified propagation speed) we obtained in Section (\ref{sectionspin0dS}), Eqn. (\ref{dSSolSpin0}). Since $V=\xi/a$, the oscillating terms decay away like their de-Sitter space counterpart. Note that this solution is valid for all wavelengths, if $A=0$.

Plugging solution (\ref{xisolution}) into Eqn. (\ref{perteqn2}), and dropping all terms that scale smaller than $k^2$, we obtain the short wavelength evolution equation for $\Phi$:
\begin{equation}
\Phi''+\left[-\frac{8\pi G_c m^2(\beta_1+\beta_2+\beta_3)(1+\alpha/\beta_1m^2)}{(1-\gamma)}+1\right]k^2\Phi=0,
\end{equation}
whose solution is a plane wave with a propagation speed of 
\begin{equation}
c_s^2{}_{\Phi}=-\frac{8\pi G_c m^2(\beta_1+\beta_2+\beta_3)(1+\alpha/\beta_1m^2)}{(1-\gamma)}+1.
\end{equation}
Setting $m^2=0$ recovers the usual solution for inflation without the presence of the vector field. 

As expected, the metric perturbation has its propagation speed modified, consistent with the argument that the ever-present vector field has changed the permeability of spacetime. Note that the conditions listed in Section (\ref{ciconditions}) mean that the propagation speed is less than unity, so $\Phi$ propagates subluminally.

\subsubsection{Long wavelength solution}

Now let us consider the evolution of long wavelength $k\eta\ll1$ perturbations. In this regime, Eqn. (\ref{perteqn1}) becomes 
\begin{equation}
\xi''=0,
\end{equation}
which has the simple solution 
\begin{equation}
\xi=D_3\eta+D_4,
\end{equation}
where $D_3$ and $D_4$ are the constants of integration for the decaying and the constant terms respectively (recall that $\eta$ runs from $-\infty$ to $0$). Plugging this back into Eqn. (\ref{perteqn2}) we get 
\begin{equation}
\Phi''+\left[2\SH-\frac{A'}{A}\right]\Phi'+\left[-2\SH^2+2\adotdota-\frac{A'}{A}\right]\Phi=0, \label{PhiLongWave}
\end{equation}
which has the following solution in an inflationary background \cite{Mukhanov:1992me}
\begin{equation}
\Phi=D_5\left(1-\frac{H}{a}\int a dt \right), \label{Phiatlongwavelength}
\end{equation}
where the cosmic time $\int dt = \int a d\eta$. At the end of inflation when the universe enters a radiation dominated era, $a\propto t^{1/2}$, so $\left(1-Ha^{-1}\int a dt\right)\approx 2/3$, thus the frozen-out spectrum of $\Phi$ is then
\begin{equation}
\Phi=\frac{2}{3}D_5. \label{endofinflationphi}
\end{equation}

The value of $D_5$ depends on the perturbation sources at the same freeze-out time, i.e. at the moment the mode leaves the Hubble radius\footnote{Technically speaking the mode freezes out when it leaves its sound horizon, i.e. at $c_s k\eta=1$.}. We will compute this value in the section below when we derive the perturbation spectra.

One important result is that, although we have an extra field on top of the inflaton, no isocurvature perturbations are produced at superhorizon scales, since Eqn. (\ref{PhiLongWave}) has no source terms on the right hand side \cite{Mukhanov:1992me}. This is an interesting departure from usual multi-scalar field models of inflation, where isocurvature perturbations are a generic feature \cite{Kofman:1985aw} (see also \cite{Mukhanov:1998fw}).

\subsection{Perturbation Spectra and the Consistency Relationship} \label{spin2spectrum} 

Observables on the CMB are ultimately sourced by the frozen-out spectra  of super-Hubble perturbation modes reentering into the Hubble radius after their sojourn outside the Hubble radius. In this Section, we compute the perturbation power spectra for both the spin-0 ``scalar'' and spin-2 ``tensor'' fields. 

We first tackle the spin-0 ``scalar'' perturbation spectrum. From Eqns. (\ref{totalEE0i}) and (\ref{spin0spaceEOM}), we replace $\xi^{(0)}{}''$ with Eqn. (\ref{xiwithphi}), and then using the non-decaying part of the short wavelength solution for $\xi^{(0)}$, Eqn. (\ref{xisolution}), we get
\begin{equation}
\frac{1}{a}(a\Phi)'=4\pi G_c \left[\frac{1+\gamma\alpha/(\beta_1m^2)}{1-\gamma}\right]\bar{\phi}'\delta \phi+Z_3\frac{k^2}{a}\int a \Phi d\eta \label{shortwavephidot}.
\end{equation}
where
\begin{equation}
Z_3\equiv \frac{(\beta_1+\beta_2+\beta_3)m^2}{\beta_1m^2(1-\gamma)}[-8\pi  G_c (\beta_1m^2+\alpha\gamma)+\gamma].
\end{equation}
Eqn. (\ref{shortwavephidot}) is the solution for $\delta \phi$; we solve the equation of motion for $\Phi$, and then insert the result into Eqn. (\ref{shortwavephidot}) to find $\delta \phi$. In the long wavelength regime $k\eta\ll0$ which we are interested in, the $Z_3$ term drops out. 

We want to match the long wavelength solution of Eqn. (\ref{shortwavephidot}) to Eqn. (\ref{Phiatlongwavelength}), which is the long wavelength solution of $\Phi$. To do that, we multiply Eqn. (\ref{Phiatlongwavelength}) with $a$ and take the time derivative of the result to get
\begin{equation}
\frac{1}{a}\left[a-\frac{\SH}{a}\int a^2 d\eta \right]'\approx\frac{4\pi  G_c \phi'^2}{a^2}\frac{a}{H}
\end{equation}
where we have used the background relation $-\SH'+\SH^2=4\pi G_c \bar{\phi}'{}^2$ and $\int a^2 d\eta =\int a dt \approx a/H$. This means that Eqn. (\ref{Phiatlongwavelength}) becomes
\begin{equation}
\frac{1}{a}(a\Phi)'=D_5 4\pi G_c \frac{\bar{\phi}'{}^2}{a^2}\frac{a}{H},
\end{equation}
which we can match to Eqn. (\ref{shortwavephidot}) to obtain the value for $D_5$
\begin{equation}
D_5=\frac{1+\gamma\alpha/(\beta_1 m^2)}{1-\gamma}\frac{a\delta \phi}{\bar{\phi}'}.
\end{equation}
Thus at the end of inflation, using Eqn. (\ref{endofinflationphi}), the metric perturbation $\Phi$ has the following value: 
\begin{equation}
\Phi=\frac{2}{3}\frac{a\delta \phi}{\bar{\phi}'}H\left[\frac{1+\gamma\alpha/(\beta_1m^2)}{1-\gamma}\right].
\end{equation}
Now using the usual result for the power spectrum of the inflaton $\delta \phi$, ${\cal{P}}_{\delta \phi}=\langle\delta \phi ^2\rangle=H^2/(2k^3)$ at large wavelengths, we obtain the spin-0 ``scalar'' power spectrum 
\begin{equation}
{\cal{P}}_{\Phi}=\langle \Phi^2\rangle=\frac{8\pi  G_c }{9}\frac{1}{\epsilon}\frac{H^2}{k^3}\left[\frac{1+\gamma\alpha/(\beta_1m^2)}{1-\gamma}\right]^2. \label{spin0power}
\end{equation}
Here the inflaton slow-roll parameter\footnote{We have used the potential-independent definition of the slow-roll parameter. See \cite{Liddle:2000cg} for a discussion on the difference between this parameter and the $\tilde{V}$ dependent definition $\epsilon=(16\pi G_*)^{-1}((\partial \tilde{V}/\partial \bar{\phi})/\tilde{V})^2$.} 
\begin{equation}
\epsilon=4\pi G_c \left(\frac{{\bar{\phi}'}}{aH}\right)^2.
\end{equation}
Comparing this to the case without the vector field (see for example \cite{DodelsonBook:2003}), the amplitude of the perturbations has been rescaled by a factor of
\begin{equation}
\frac{{\cal{P}}_{\Phi}}{\bar{{\cal{P}}}_{\Phi}}\equiv\frac{ G_c }{G_*}\left[\frac{1+\gamma\alpha/(\beta_1m^2)}{1-\gamma}\right]^2; \label{spin0factor}
\end{equation}
where $\bar{{\cal{P}}}_{\Phi}$ is the power spectrum without the vector field. One can check that by setting $m^2=0$, we recover the usual relation for the power spectrum (\ref{spin0power}) where ${\cal{P}}_{\Phi}/\bar{{\cal{P}}}_{\Phi}=1$. Note that the index of the power spectrum does not change since vector field background precisely tracks that of the inflaton.

As an aside, we comment on the use of the relation $\delta \phi \sim H$ above. One can argue that there is no reason to believe that the inflaton perturbation at small scales will evolve similarly to the usual no-vector-field case. However, since there is no direct coupling between the inflaton and the vector field, any effects that the inflaton perturbation will experience due to the vector field has to be through the metric perturbation which is at most a second order effect.

%We can understand this rescaling (\ref{spin0factor}) by noting that, like its background counterpart, the vector perturbation tracks the behaviour of the inflaton perturbation, and effectively changes the amplitude of the metric perturbation.

This rescaling of the spin-0 amplitude is degenerate with the slow-roll parameter $\epsilon$, thus from the standpoint of the spin-0 perturbation alone we cannot disentangle them. Nevertheless, there is hope for us to distinguish between them if we can observe the gravitational wave spectrum, as we will see later. 

From Eqn. (\ref{dSSpin2Sol}) and following \cite{DodelsonBook:2003}, we can then compute the power spectrum for spin-2 ``tensor'' perturbations to be
\begin{equation}
{\cal{P}}_{H_T^{(2)}}=8\pi G_*\sqrt{1-\gamma}\frac{H^2}{k^3}. \label{spin2power}
\end{equation}
Since for non-superluminal propagation of the gravity waves we need $\gamma<0$, the spin-2 ``tensor'' power spectrum is boosted, making the tensor spectrum possibly more accessible to observations (currently we have no detection of a spin-2 spectrum \cite{Peiris:2003ff}). Of course, this value is degenerate with the Hubble parameter $H$ so measurement of the spin-2 spectrum does not allow us to put a constraint on the LV vector field parameters.

If we assume natural ${\cal{O}}(-1)$ values for $\beta_1+\beta_3$, then $\gamma\approx (m/m_p)^2$ where $m_p^2=1/(8\pi G_N)$. Thus if $m$ is of order $m_p$ which is to say that Lorentz-violating effects are a symptom of new Planck scale physics, then there will be a boost in our chances at observability of gravity waves in the CMB. However, as shown in \cite{Carroll:2004ai}, primordial nucleosynthesis constraints $m$ to be an order of magnitude smaller than $m_p$. 

Before we close our investigation of LV vector fields in an inflationary universe, we note that the presence of this vector field has violated the so-called inflationary consistency relation. One can show that any single slow-roll scalar field model of inflation with slow-roll parameter $\epsilon$, the spectral index for the spin-2 spectrum $n_T=-2\epsilon$. But the ratio of the amplitudes for the spin-2 and spin-0 spectra is proportional to $\epsilon$, so this relation states that  
\begin{equation}
\frac{{\cal{P}}_{H_T^{(2)}}}{{\cal{P}}_{\Phi}}=-\frac{9}{2}n_T.
\end{equation}

But here, with the vector field permeating the universe, from Eqns. (\ref{spin0power}) and (\ref{spin2power}) we obtain the modified relation
\begin{equation}
\frac{{\cal{P}}_{H_T^{(2)}}}{{\cal{P}}_{\Phi}}=-\frac{9}{2}\frac{(1+8\pi G\alpha)(1-\gamma)^{3/2}}{[1+\gamma\alpha/(\beta_1m^2)]^2}n_T.
\end{equation}
In other words, if we believe in the single slow-roll scalar field model of inflation, then measuring $n_T$ will allow us to put a constraint on the parameters of the theory. 

\section{Conclusions} \label{Conclusion}

We have shown that perturbations of a fixed-norm, timelike vector field have a consistent quantum theory. If we insist on positivity of the Hamiltonian and non-tachyonic propagation of the quantized field, then we find that $\beta_1>0$ for the former and $\beta_1+\beta_2+\beta_3\geq0$ for the latter. In addition, non-superluminal propagation of the spin-0 field and gravity waves requires that $(\beta_1+\beta_2+\beta_3)/\beta_1\leq1$ and $\beta_1+\beta_3\leq0$ respectively. These facts imply that the sign of the background energy density of the vector field is non-positive. 

We then considered the evolution of the vector perturbations in a FRW universe. Specializing to a de-Sitter spacetime, we find that the perturbations decay away exponentially. We argued that this is because there is no other matter perturbations for the vector field to track.

Following that, we investigated its evolution in the presence of an inflaton field. As expected, the vector perturbation is now sourced by the inflaton perturbation, making it non-decaying and thus imparting non-trivial effects on the metric perturbations. We derived the generated primordial spectrum for both spin-0 (approximately) and spin-2 (exactly) perturbations, and showed that their amplitude is rescaled. In particular, we noted that if the norm $m$ of the vector field is large enough, it might lead to a boosted spectrum of B-mode polarization in CMB. We also showed that the presence of the vector perturbations also violates the inflationary consistency relation.

Finally, the ultimate test of the observability of LV vector field induced modifications in the CMB lies in our ability to compute the evolution of perturbations from Hubble reentry through to the surface of last scattering. This entails solving the multicomponent Boltzmann evolution equations, adding the vector perturbation on top of the usual photon-baryon-dark matter fluid, an endeavour that can only be done numerically. 
 
\section{Acknowledgments}

The author would like to thank Christian Armendariz-Picon, Jim Chisholm, Scott Dodelson, Vikram Duvvuri, Christopher Eling, Christopher Gordon, Stefan Hollands, Wayne Hu, Ted Jacobson, Rocky Kolb, Alan Kostelecky, David Mattingly, Takemi Okamoto, Clem Pryke, Eduardo Rozo, Robert Wald and especially Sean Carroll for useful and insightful discussions. The author is supported by the David and Lucille Packard Foundation. This research was carried out at the University of Chicago, Kavli Institute for Cosmological Physics and was supported (in part) by grant NSF PHY-0114422. KICP is a NSF Physics Frontier Center
\appendix

\section{Eigenmode Decomposition} \label{AppA}

In this paper, we have chosen to work in eigenmodes of the irreducible representations of SO(3). This formalism was first applied to cosmological perturbations by Bardeen \cite{Bardeen:1980kt}. We present a short review in this appendix for completeness and also to pin down conventions, specializing to a spatially flat universe. For a good comprehensive review, see \cite{Kodama:1984bj}.

In a background FRW spacetime that is foliated into spatial hypersurfaces labeled by cosmic time, the linear perturbations are just functions living in a spatially SO(3) and translationally invariant background. If further, there are no background quantities that violate this symmetry (e.g. a background vector field with spatial components or an inhomogenous scalar field), the perturbation variables decomposed using these eigenmodes are decoupled from each other. 

The spin-0, spin-1 and spin-2 orthonormal modes are eigenmodes of the Laplace-Beltrami operator
\begin{eqnarray}
\partial^2 Y^{(0)}+k^2Y^{(0)}&=&0 \\
\partial^2 Y_i^{(\pm 1)}+k^2Y_i^{(\pm 1)}&=&0 \\
\partial^2 Y_{ij}^{(\pm 2)}+k^2Y_{ij}^{(\pm 2)}&=&0.
\end{eqnarray}
These modes obey the following conditions
\begin{equation}
\partial_i Y^i{}^{(\pm 1)}=0~,~\partial_i Y^{ij}{}^{(\pm 2)}~,~Y_i^i{}^{(\pm 2)}=0
\end{equation}
i.e. spin-1 modes are divergenceless, and the spin-2 modes are traceless and transverse. With these conditions, we can derive their explicit form
\begin{eqnarray}
Y_{\vec{\mathbf{k}}}^{(0)}(\vec{\mathbf{x}})&=&e^{i\vec{\mathbf{k}}\cdot\vec{\mathbf{x}}} \\
Y_{\vec{\mathbf{k}}}^{i}{}^{(\pm 1)}(\vec{\mathbf{x}})&=&\frac{\epsilon^{ijp}}{\sqrt{2}k}(k_ln_p\pm i k_pn_l)e^{i\vec{\mathbf{k}}\cdot\vec{\mathbf{x}}}\\
Y_{\vec{\mathbf{k}}}^{ij}{}^{(\pm 2)}(\vec{\mathbf{x}})&=&-\sqrt{\frac{3}{8}}k^{-2}[\epsilon^{ilp}(k_ln_p\pm i k_pn_l)][\epsilon^{jrs}(k_rn_s\pm i k_sn_r)]e^{i\vec{\mathbf{k}}\cdot\vec{\mathbf{x}}}
\end{eqnarray}
where $\vec{\mathbf{n}}$ is a unit vector such that $\vec{\mathbf{k}}\cdot\vec{\mathbf{n}}=0$, $\epsilon^{ijk}$ is the Levi-Civita symbol and $k=|\vec{\mathbf{k}}|$. Here we have restored the $\vec{\mathbf{k}}$ subscript and the function argument $(\vec{\mathbf{x}})$ to the mode function. 

Using these basis modes, we construct other modes. Following the convention of \cite{Kodama:1984bj},
\begin{eqnarray}
Y_i^{(0)}&=&-k^{-1}\partial_i Y^{(0)}\\
Y_{ij}^{(0)}&=&\left(k^{-2}\partial_i \partial_j+\frac{1}{3}\gamma_{ij}\right)Y^{(0)}\\
Y_{ij}^{(\pm 1)}&=&-\frac{1}{2k}(\partial_i Y_j^{(\pm 1)}+\partial_j Y_i^{(\pm 1)}).
\end{eqnarray}
In particular the following identities are especially useful when deriving the commutation relations and the Hamiltonian in Section (\ref{QuantumFlat})
\begin{eqnarray}
Y_i{}^{(0)}_{\vec{\mathbf{k}}}&=&-Y_i{}^{(0)}_{-\vec{\mathbf{k}}} \\
Y_i{}^{(\pm 1)}_{\vec{\mathbf{k}}}&=&-Y_i{}^{(\pm 1)}_{-\vec{\mathbf{k}}}.
\end{eqnarray}

We end this appendix with a warning that different authors often use different conventions for the explicit mode functions. In this work, the explicit mode functions never appear.

%\section{Choosing a Gauge} \label{AppB}

\section{Equations of Motion in Real Space} \label{AppC}

In this appendix, we collect the real space equations for reference.

The perturbed component of the equation of motion Eqn. (\ref{EOMforu}) is
\begin{eqnarray}
&&-6(\beta_1+2\beta_2+\beta_3)\SH^2\Phi+6\beta_2\adotdota\Phi \nonumber \\
&&+\SH\left[(2\beta_1+\beta_2)\partial_i V^i+\beta_3(\partial_i V^i+\partial_i B^i)+3\beta_2\Phi'+3(2\beta_1+\beta_2+2\beta_3)\Psi'\right] \nonumber \\
&&-\beta_3(\partial_i\partial^i\Phi-\partial_iB'{}^i+\partial_iV'{}^i)-\beta_2(\partial_i V'{}^i+3\Psi'')+a^2\delta \lambda=0 
\end{eqnarray}
for the $\nu=0$ component and
\begin{eqnarray}
&&-\left[2\frac{\alpha}{m^2}\SH^2-\frac{\alpha}{m^2}\adotdota+\beta_1\adotdota\right](B_i-V_i) \nonumber \\
&&+\SH\left[(\beta_1+\frac{\alpha}{m^2})\partial_i \Phi+2\beta_1(V_i'-B_i')\right] \nonumber \\
&&+\frac{1}{2}(-\beta_3+\beta_1)\partial_{[j}\partial^j B_{i]}-\beta_1\partial_j\partial^j V_i-(\beta_2+\beta_3)\partial_i\partial^j V_j \nonumber \\
&&-(\beta_1+\beta_3)\partial_j h'{}^j{}_i+\beta_1\partial_i\Phi'-\frac{\alpha}{m^2}\partial_i \Psi'-\beta_1(B_i''-V_i'')=0
\end{eqnarray}
for the $\nu=i$ component. 

The stress-energy tensor has the following components:
\begin{eqnarray}
\delta T^0{}_0&=&2\frac{m^2}{a^2}\left\{-3(\beta_1+3\beta_2+\beta_3)\SH^2\Phi+\beta_1a^{-1}\left[a\partial_i(B^i-V^i)\right]'\right.\nonumber \\
&&\left.+(\beta_1+3\beta_2+\beta_3)\SH(\partial_i V^i+3\Psi')+\beta_1\partial^2\Phi\right\} \\
\delta T^0{}_i&=&2\frac{m^2}{a^2}\left\{\left[-2(\beta_1+3\beta_2+\beta_3)\SH^2+(3\beta_2+\beta_3)\adotdota\right](V_i-B_i) \right. \nonumber \\
&&-\beta_1a^{-2}\left[a^2(V_i'-B_i')\right]'-\beta_1a^{-1}(a\partial_i\Phi)' \nonumber \\
&& \left. +\frac{1}{2}(-\beta_1+\beta_3)\left[\partial^2(B_i-V_i)-\partial_i\partial^j(B_j-V_j)\right]\right\},\\
\delta T^i{}_j&=&2\frac{m^2}{a^2}\left\{(\beta_1+3\beta_2+\beta_3)\left[\SH^2-2\adotdota\right]\Phi\delta^i_j-(\beta_1+3\beta_2+\beta_3)\SH\Phi'\delta^i_j \right. \nonumber \\
&& +a^{-2}\left[a^2(\beta_2\partial_k V^k\delta^i_j+(\beta_1+3\beta_2+\beta_3)\Psi'\delta^i_j \right. \nonumber \\
&&\left.\left.+\frac{1}{2}(\beta_1+\beta_3)(\partial_jV^i+\partial^iV_j+2h^i{}_j')\right]'\right\}
\end{eqnarray}

\bibliography{vecpertnot}

\end{document}